\documentclass[preprint,10pt,plainnat]{aastex6}
\usepackage{amsmath,amstext}
\usepackage[figure,figure*]{hypcap}
\usepackage{newtxmath} 

\shorttitle{Quantitative indexing and Tardigrade analysis of exoplanets}
\shortauthors{J. M. Kashyap}

\begin{document}

\title{Quantitative indexing and Tardigrade analysis of exoplanets}
\author{J. M. Kashyap\altaffilmark{1}}
\email{kas7890.astro@gmail.com}
\altaffiltext{1}{Department of Physics, Jyoti Nivas College, Bengaluru-560095, Karnataka, India}

\begin{abstract}
Search of life elsewhere in the galaxy is very fascinating area for planetary scientists and astrobiologists. Earth Similarity Index (ESI) is defined as geometrical mean of four physical parameters (Such as radius, density, escape velocity and surface temperature), which is ranging from 1 (identical to Earth) to 0 (dissimilar to Earth). In this work, ESI is re-defined as six parameters by introducing the two new physical parameters like revolution and surface gravity and is called as New Earth Similarity Index (NESI).
The main focus of this paper is to search Tardigrade water-life on exoplanets by  varying the temperature parameter in NESI, which is called as Tardigrade Similarity Index (TSI), which is ranging from 1 (Tardigrade can survive) to 0 (Tardigrade Cannot survive). Here the NESI and TSI is cataloged and analyzed for almost 3370 confirmed exoplanets.
\end{abstract}


\section{Introduction}

 Exploring the unknown worlds outside our solar system (i.e., exoplanets) is the new era of the current research. Presently with the huge flow of data from Planetary Habitability Laboratory PHL-HEC  \footnote{http://phl.upr.edu/projects/habitable-exoplanets-catalog/data/database}, maintained by university of Puerto Rico, Arecibo. Indexing will be a main criteria to give a proper structure to these raw data from space missions such as CoRoT and Kepler. Nearly half a decade ago, Schulze-Makuch et al.(2011) has defined Earth Similarity Index (ESI) as a geometrical mean of four physical parameters (such as: radius, density, escape velocity and surface temperature). In this paper, the New Earth Similarity Index (NESI) is re-defined as geometrical mean of six physical parameters (namely: radius, density, escape velocity, surface temperature, revolution, surface gravity). Since revolution is not directly available as the raw data, here the values are calculated for 3370 (as of September 2016) confirmed exoplanets.
We are always interested in life-forms, which can survive outside our planet and tardigrade is the likely candidate \citep{Ing}. Tardigrade primarily known as moss piglets or water bears \citep{c}, and they can survive at different temperature scales (Examples: $ 151 \deg C$ for few minutes \citep{H}, $-20 \deg C$ for $30$ years \citep{T}, $-200 \deg C$ for days \citep{H}, $-272 \deg C$ for few minutes \citep{B} in lower temperature scale).\\
The structure of the paper is as follows: In section 2, the NESI has been introduced and analysis of 3370 confirmed exoplanets are done, section 3 has the results of TSI, and section 4 gives the discussion and conclusion part of the work.

\section{New Earth Similarity Index and its analysis}

In 2011, Schulze-Makuch et al., defined the Earth similarity index as
\begin{equation}
ESI_x = {\left[1-\Big|
\frac{x-x_0}{x+x_0}\Big| \right]^{w_x}}\,,
\label{eq:esi}
\end{equation}
where $x$ is the planetary property of the exoplanet, $W_X$ is the weight exponent and $x_0$ is the reference to Earth in ESI.\\

In this paper we follow the above equation to calculate the NESI of individual planetary property, with the weight exponents as defined below. The New Earth Similarity Index (NESI) is defined as the geometrical mean of six physical parameters (such as: radius, density, escape velocity, surface gravity, revolution and surface temperature). Mathematically it can be denoted as: 

\begin{equation}
NESI = ({{NESI_R} \times {NESI_\rho}} \times {NESI_T \times NESI_{V_e}} \times {NESI_{Re} \times NESI_{g}})^{1/6}
\end{equation}

The weight exponents for upper and lower limits of parameters are calculated as (Schulze-Makuch et al.(2011)): radius 0.5 to 1.9 EU, mass 0.1 to 10 EU, density 0.7 to 1.5 EU, surface temperature 273 to 323 K and escape velocity 0.4 to 1.4 EU. Similarly, we define the limits of gravity as 0.16 to 17 EU and revolution as 0.61 to 1.88 EU.\\

{\it Gravity and Revolution}:\\
The human centrifuge experiment clearly showed the untrained humans can tolerate 17 EU, with eye balls in \citep{Br}. The revolution is scaled on the basis of habitable zone of sun-like stars.

\begin{table}[h!]
\begin{center}
\caption{NESI Parametric Table}\label{table:2.1}
\begin{tabular}{l*{4}{c}r}
\hline
Planetary Property &Ref. Value & Weight Exponents\\
            & for NESI  & for NESI & \\ \hline
Mean Radius	& 1EU  & 0.57 \\ 
Bulk Density &	1EU  &	1.07 \\ 
Escape Velocity	&1EU&	0.70\\ 
Revolution & 1EU & 0.70\\
Surface gravity& 1EU & 0.13\\
Surface Temperature	& 288K&5.58\\
\hline
\end{tabular}
\end{center}
\end{table}
Here, EU = Earth Units, where Earth’s radius is 6371 km, density is 5.51 g/cm$^3$, escape velocity is 11.19 km/s, Revolution is 365.25 days and surface gravity is 9.8 m/s$^2$.

The NESI is further divided into Interior NESI and Surface NESI. The interior NESI is defined as the geometrical mean of radius and density.
Mathematically it takes the form:

\begin{equation}
NESI_I = ({{NESI_R} \times {NESI_\rho}})^{1/2}\,,
\end{equation}

Similarly, the surface NESI is defined as the geometrical mean of surface temperature, escape velocity, revolution of the planet and surface gravity. Empirically is of the form:
\begin{equation}
NESI_S = ({NESI_T \times NESI_{V_e}} \times {NESI_{Re} \times NESI_{g}})^{1/4}
\end{equation}

The global NESI takes the form:

\begin{equation}
NESI = ({{NESI_I} \times {NESI_S}})^{1/2}\,,
\end{equation}

\begin{table}[h!]
\centering
\caption{A sample of calculated NESI}
\begin{tabular}{l*{8}{c}r}
\hline
Names & Radius & Density &Temp & E. Vel  & Rev & g &$NESI_S$ & $NESI_I$ & NESI\\ 
\ &  (EU) & (EU) & (K) & (EU) &  \ & \ &  \\\hline
Earth &	1.00 &	1.00 &	288	&1.00	&	1.00 & 1.00 &1.00 &1.00	&1.00\\
Mars &	0.53 &	0.73 &	240	&0.45	&1.88& 0.37 &	0.74	&0.81	&0.77\\
Kepler-438 b & 1.12 & 0.90  & 312 & 1.06& 0.09& 1.01& 0.69 &0.96 & 0.81\\
GJ 667C c & 1.54 & 1.05 & 280 & 1.57& 0.07& 1.61& 0.66  & 0.92 & 0.78\\
Kepler-296 e	& 1.48	& 1.03	 & 303 & 1.50 & 0.08& 1.52& 0.67 & 0.93 & 0.79\\
\hline
\end{tabular}
\label{Table:2.1}
\end{table}

The entire analysis for 3370 exoplanets is cataloged in separate rocky and gas exoplanets files, which is available at\citep{Madhu1}.

\begin{figure}[h!]
\centering        
\includegraphics[width=13cm,angle=0]{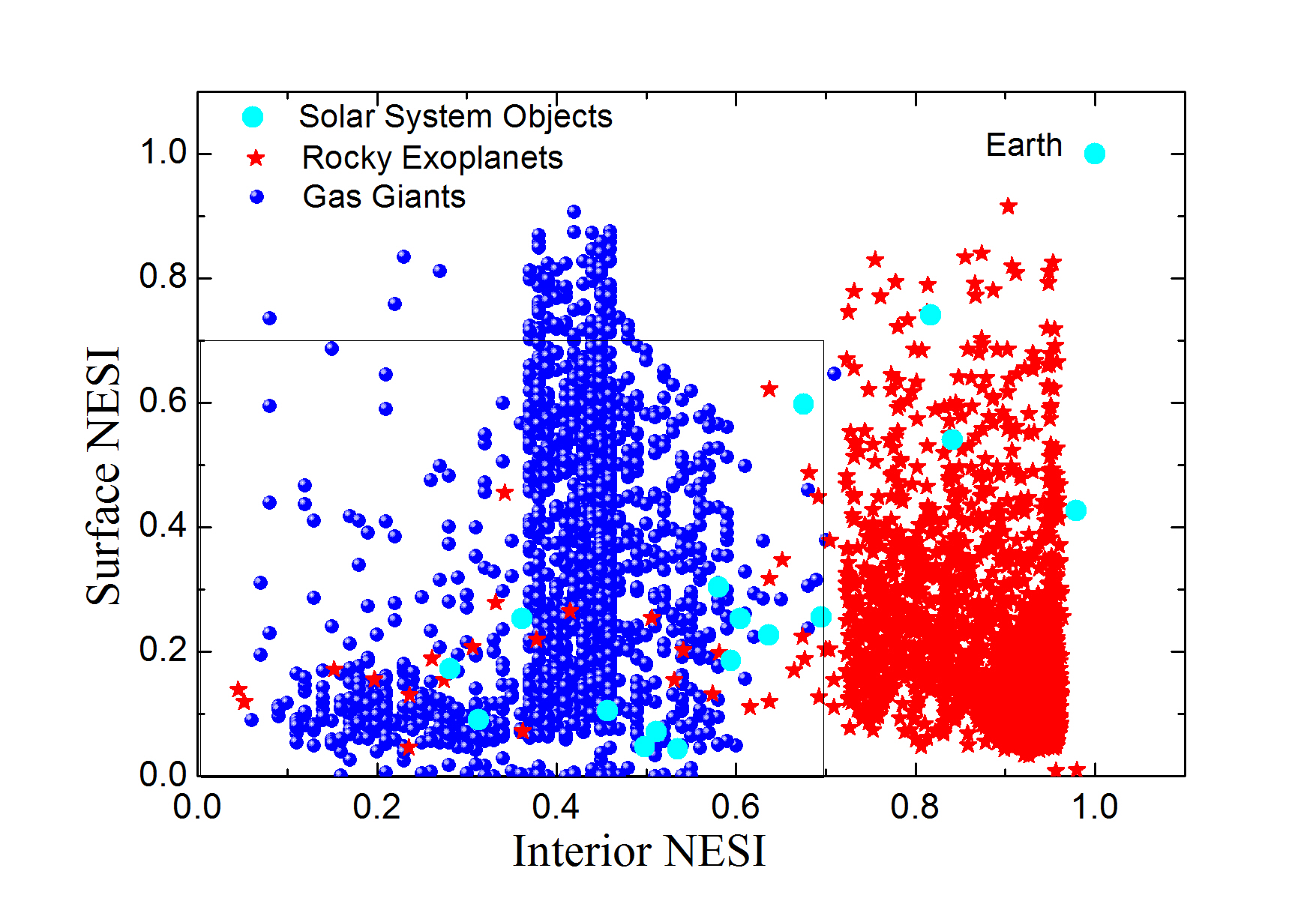} 
\caption{Plot of interior NESI versus surface NESI. Blue dots are the giant planets, red dots are the rocky planets, and cyan circles are the Solar System objects (Table~2). However, the limit for NESI is marked by a solid line which is $\sim 0.70$.}
\label{fig:ESIplot} 
\end{figure}
\newpage 
The peaks in the plot are due to distribution phenomenon of similarity indices. This scatter plot denotes the optimistic range of each and every gas giants and rocky exoplanets, with respect to Earth-like planets.

\section{Tardigrade Similarity Index}

The Tardigrade Similarity Index (TSI) is defined similarly as NESI, with different weight exponent for surface temperature. Mathematically it can be denoted as:  


\begin{equation}
TSI = ({{TSI_R} \times {TSI_\rho}} \times {TSI_T \times TSI_{V_e}} \times {TSI_{Re} \times TSI_{g}})^{1/6}
\end{equation}

The weight exponent range for all the physical parameters are same as NESI, except the temperature parameter, which ranges from 1.15 to 424.15 K for tardigrade to survive.

\begin{table}[h!]
\begin{center}
\caption{TSI Parametric Table}\label{table:3.1}
\begin{tabular}{l*{4}{c}r}
\hline
Planetary Property &Ref. Value & Weight Exponents\\
            & for TSI  & for TSI & \\ \hline
Mean Radius	& 1EU  & 0.57 \\ 
Bulk Density &	1EU  &	1.07 \\ 
Escape Velocity	&1EU&	0.70\\ 
Revolution & 1EU & 0.70\\
Surface gravity& 1EU & 0.13\\
Surface Temperature	& 288K & 0.21\\
\hline
\end{tabular}
\end{center}
\end{table}
Here, EU = Earth Units, where Earth’s radius is 6371 km, density is 5.51 g/cm$^3$, escape velocity is 11.19 km/s, Revolution is 365.25 days, and surface gravity is 9.8 m/s$^2$.\\ 

The TSI is classified into Interior TSI and Surface TSI. The interior TSI is the geometrical mean of TSI radius and density.

\begin{equation}
TSI_I = ({{TSI_R} \times {TSI_\rho}})^{1/2}\,,
\end{equation}\\

Similarly, the surface TSI is defined as the geometrical mean TSI of surface temperature, escape velocity, revolution of the planet and surface gravity.

\begin{equation}
TSI_S = ({TSI_T \times TSI_{V_e}} \times {TSI_{Re} \times TSI_{g}})^{1/4}
\end{equation}

The global TSI takes the form:

\begin{equation}
TSI = ({{TSI_I} \times {TSI_S}})^{1/2}\,,
\end{equation}

\begin{table}[h!]
\centering
\caption{A sample of calculated TSI}
\begin{tabular}{l*{8}{c}r}
\hline
Names & Radius & Density &Temp & E. Vel  & Rev & g &$TSI_S$ & $TSI_I$ & TSI\\ 
\ &  (EU) & (EU) & (K) & (EU) &  \ & \ &  \\\hline
Earth &	1.00 &	1.00 &	288	&1.00	&	1.00 & 1.00 &1.00 &1.00	&1.00\\
Mars &	0.53 &	0.73 &	240	&0.45	&1.88& 0.37 &	0.84 &0.81	&0.82\\
Kepler-438 b & 1.12 & 0.90  & 312 & 1.06& 0.09& 1.01& 0.73 &0.96 & 0.84\\
GJ 667C c & 1.54 & 1.05 & 280 & 1.57& 0.07& 1.61& 0.67  & 0.92 & 0.78\\
Kepler-296 e	& 1.48	& 1.03	 & 303 & 1.50 & 0.08& 1.52& 0.69 & 0.93 & 0.80\\
\hline
\end{tabular}
\label{Table:3.2}
\end{table}

\begin{figure}[h!]
\centering        
\includegraphics[width=13cm,angle=0]{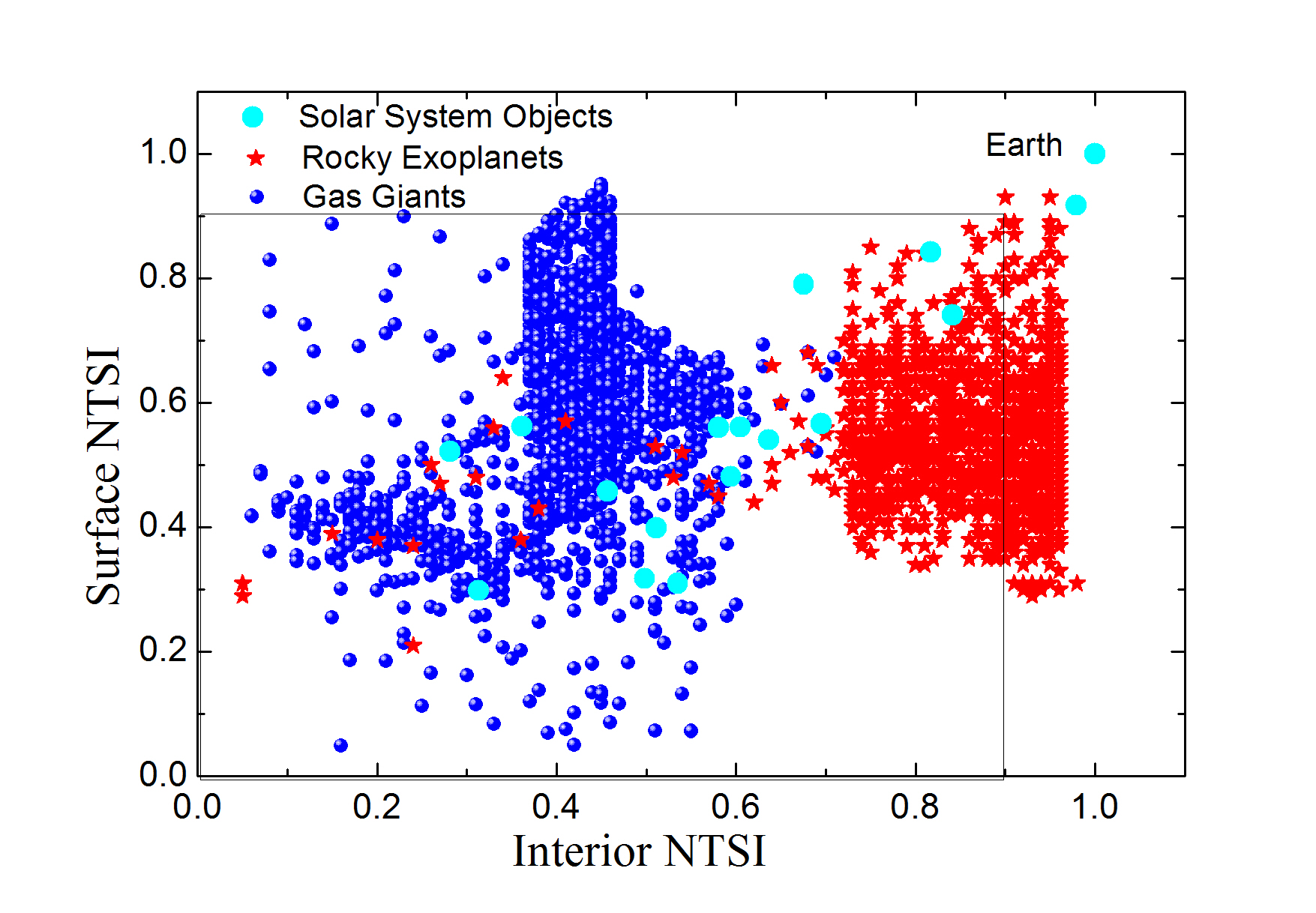} 
\caption{Plot of interior NTSI versus surface NTSI. Blue dots are the giant planets, red dots are the rocky planets, and cyan circles are the Solar System objects (Table~4). However, the optimistic limit for NTSI is marked by a solid line which is $\sim 0.90$.}
\label{fig:TSIplot}
\end{figure}

Similar to NESI, the peaks in the plot gives the distribution phenomenon of similarity indices. TSI scatter plot denotes the optimistic range of each and every gas giants and rocky exoplanets, with respect to Tardigrade survival ability. 

\section{Discussion and Conclusion}
The Search for Earth-twin is becoming more vibrant area in planetary science research. We know that, ESI was defined for only four physical parameters. But it is necessary to understand the need for compiling more physical parameters to find Earth-like planet. Hence, in this paper the NESI has been introduced to get more accurate results.The results obtained above clarifies it on comparison with ESI with 4 parameters \citep{s, Madhu}.   
In 2008 \citep{Ing}, showed that Tardigrade could survive in space for 10 days.
Thus the Tardigrade Similarity Index (TSI) is introduced and analyzed for all 3370 exoplanets, where this extremophile life could survive.
The future work in this area for the quest in search of life can be done by upgrading NESI with more physical parameters(such as: tilt of the planet, albedo factor, magnetic field, ... etc). Finally, in order to sustain life atmospheric study in detail plays a major role, therefore analyzing the astrochemistry of the exoplanets is very crucial.

\section*{Acknowledgments}
This research has made use of the Extrasolar Planets Encyclopaedia at {\tt http://www.exoplanet.eu}, Exoplanets Data Explorer at {\tt http://exoplanets.org}, the Habitable Zone Gallery at {\tt http://www.hzgallery.org/}, the NASA Exoplanet Archive, which is operated by the California Institute of Technology, under contract with the National Aeronautics and Space Administration under the Exoplanet Exploration Program at\\
{\tt http://exoplanetarchive.ipac.caltech.edu/} and 
NASA Exoplanet Archive at\\
{\tt http://exoplanetarchive.ipac.caltech.edu} and NASA Astrophysics Data System Abstract Service.

%
%
\end{document}